УДК 004.9:66.013.512

В.В. Мигунов, Р.Р. Кафиятуллов

# ЗАЩИТА ИНФОРМАЦИИ В КОМПЛЕКСНОЙ САПР РЕКОНСТРУКЦИИ ПРОМЫШЛЕННЫХ ПРЕДПРИЯТИЙ

**Введение**. По мере перехода к электронному представлению информации, которое делает возможным ее хранение и транспортировку на все более компактных носителях и передачу со все большими скоростями по каналам связи, задача защиты информации (ЗИ) в электронном виде приобретает особую актуальность. Возникающие здесь проблемы и методы очень разнообразны: от законодательных, административных и организационных до таких аспектов технической защиты, как предотвращение внедрения программ-вирусов или физическая защита помещений (п. 5.1.3. [1]).

Теория и практика ЗИ в нашей стране развиваются как путем распространения глубоко разработанных методов защиты сведений, составляющих государственную тайну, на конфиденциальные сведения [1], так и на основе относительно новых разработок в сфере защиты коммерческой тайны [2, 3]. При этом в первую очередь рассматривается ЗИ от умышленных действий. Например, в п. 5.1.2. [1] говорится: "Основными направлениями защиты информации являются: обеспечение защиты информации от хищения, утраты, утечки, уничтожения, искажения и подделки за счет НСД и специальных воздействий; обеспечение защиты информации от утечки по техническим каналам при ее обработке, хранении и передаче по каналам связи" (НСД - несанкционированный доступ). В [3]: "В настоящей работе предпринята попытка возможно более полного охвата угроз безопасности субъектов информационных отношений. Однако следует иметь в виду, что научно-технический прогресс может привести к появлению принципиально новых видов угроз и что изощренный ум злоумышленника способен придумать новые способы преодоления систем безопасности, НСД к данным и дезорганизации работы информационных систем". Лишь во вторую очередь принято рассматривать угрозы от неумышленных действия. В частности, для САПР эти угрозы и методы борьбы с ними изучены мало.

В противоположность подходам к ЗИ, акцентированным на противодействии злому умыслу, настоящая работа посвящена изучению угроз информации, возникающих вследствие непредумышленных действий пользователей САПР, и методов ЗИ, ограниченных сферой влияния программного обеспечения САПР. В основе рассмотрения лежит практический опыт развития комплексной САПР реконструкции предприятий TechnoCAD GlassX, эксплуатируемой в проектно-конструкторском подразделении крупного предприятия химической промышленности [4] в течение 10 лет. Из особенностей такой САПР, охарактеризованных в [5], существенной для ЗИ явилась комплексность. Отметим, что именно непредумышленные действия являются основными угрозами информации. По сведениям из [2]: "... в 1998 году. Основные причины повреждений электронной информации распределились следующим образом: неумышленная ошибка человека - 52% случаев, умышленные действия человека - 10% случаев, отказ техники - 10% случаев, повреждения в результате пожара - 15% случаев, повреждения водой - 10% случаев". В [3] приводятся данные, относящиеся к более позднему времени: "Согласно статистики, 65% потерь - следствие непреднамеренных ошибок. Пожары и наводнения можно считать пустяками по сравнению с безграмотностью и расхлябанностью".

**Угрозы информации в сфере влияния САПР.** В первую очередь беспокоящие проектировщика угрозы информации в САПР – это угрозы порчи или утраты электронных документов (файлов с чертежами), приводящие к необходимости повторно выполнять уже сделанную работу. По опыту эксплуатации САПР выделены 4 угрозы, которые в первую очередь отмечаются пользователями:

1. Случайное удаление, изменение элементов чертежа.

2. Утеря результатов работы за период с последнего сохранения чертежа на диск вследствие отказов: сети электроснабжения, узлов компьютера, операционной системы и самой САПР.

3. Случайная подмена чертежа другим с тем же именем файла. Для идентификации чертежей проектировщики часто употребляют номер проектируемого корпуса. При этом монтажный чертеж трубопровода и схема автоматизации, созданные на разных рабочих местах, могут иметь одно и то же имя файла. При передаче чертежей между рабочими местами возникает угроза неумышленной замены чертежа.

4. Утрата уверенности в целостности чертежа, возникающая при отсутствии системы ответственного электронного архивирования (хранения) файлов чертежей. Чертеж может измениться в результате неосторожных действий при его просмотре или печати, при использовании его как заготовки для создания другого чертежа.

Реализация ЗИ в САПР от этих угроз далее описана соответственно как "Откат", "Автосохранение", "Автоматическое резервное копирование" и "Электронная подпись".

**Откат.** Откат, или возврат изменений (команды "Отменить" и "Вернуть") - широко применяемый метод борьбы с 1й угрозой. Его теория глубоко развита для баз данных [6] как работа с транзакциями. В условиях комплексной САПР возникает специфика, связанная с тем, что при выходе на довольно продолжительное время в режим работы в специализированном проблемно-ориентированном расширении [7] пользователь занимается параметрическим проектированием, не имеет связи с чертежом и накапливает информацию не в нем, а в создаваемом параметрическом представлении (ПП). Примеры таких расширений: проектирование профилей наружных сетей водоснабжения и канализации, молниезащиты зданий и сооружений. Функции отката рассмотрим сначала для чертежа, затем для ПП его частей.

Все операции разработки чертежа удается свести к двум первичным, атомарным - добавление и удаление элемента чертежа. Изменение типа линии, слоя, цвета и др. рассматриваются как удаление и последующее добавление в чертеж измененного элемента. Операции над множеством элементов (сдвиг, растяжение и др.) моделируются как множество одинаковых операций, каждая из которых выполняется над одним элементом. Вводится понятие "шаг изменения", которое включает все операции, проведенные пользователем над однократно выбранными элементами. Тем самым достигается соответствие интуитивного представления об "изменении" и количества возвращаемых за один шаг атомарных изменений - аналог транзакций в базах данных. На рис.1 каждый шаг обозначен параллелограммом, "стабильные" состояния чертежа - флажками, а стрелки показывают направление изменений. Последовательность шагов при возврате изменений существенна, иначе: может потребоваться удаление элементов, которых в чертеже нет; пользователь не будет знать, какое изменение возвратится в первую очередь. Учитываются изменения чертежа в последнем сеансе работы с ним, возврат возможен до завершения сеанса.

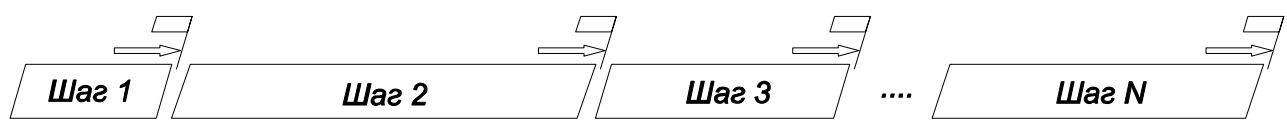

*Рис.1. Модель пошаговых изменений - транзакций*

В рабочий файл последовательно записываются удаляемые и добавляемые геометрические элементы вместе с признаком - добавлен или удален элемент. В результате, как показано на рис.2, один шаг представляется в виде сцепленных в железнодорожный состав различных геометрических элементов, образующих как бы сумму атомарных изменений. Внутри одного шага последовательность элементов несущественна, если все элементы разные. Если же для удобства программирования приходится в один шаг включать, например, удаление и последующее восстановление одного и того же элемента, то порядок следования атомарных изменений при возврате должен быть строго обратным.

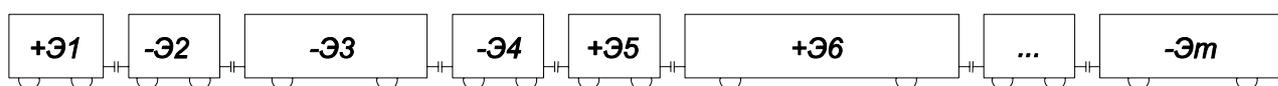

*Рис.2. Общая структура одного шага изменения чертежа*

Для возврата изменения делается проход по файлу, пока номер изменения совпадает с нужным. Для каждого из найденных элементов с этим номером изменения производится следующее: если он был удален, он добавляется в чертеж; если он был добавлен, то в чертеже ищется совпадающий с ним и удаляется из чертежа. Применяя команды "Отменить" и "Вернуть", можно как угодно долго перебирать состояния чертежа в текущем сеансе работы.

ПП части чертежа есть комплекс ее свойств, или информационная модель. Видимая геометрическая часть генерируется по ПП. С точки зрения отката наиболее существенное отличие работы с ПП от работы непосредственно с чертежом заключается в интеллектуальных связях элементов друг с другом. При удалении одной трубы из ПП аксонометрической схемы трубопроводной системы автоматически удаляются все нанесенные на нее обозначения трубопроводной арматуры, элементов трубопроводов, тексты и сноски, указывающие на трубу и удаленные обозначения, выносные линии размеров.

В этих условиях система отката не имеет явно очерченных границ того, что же изменяется за один шаг. Верхней границей является ПП полностью. В рамках модульной технологии разработки расширений САПР все ПП преобразуются к единой компактной форме, задаваемой адресом в памяти и длиной [7], поэтому заменяемым объектом выбрано все ПП. Это решение универсально для различных расширений САПР.

При любом изменении новая версия ПП записывается в рабочий файл. Операции над множеством здесь возможны, но их не надо выражать в множестве актов удаления и добавления, каждому шагу изменений соответствует одно ПП. Отмена изменений и возврат реализуются путем перехода к предыдущей или к следующей версии ПП с полной его заменой. Ситуации соответствует рис.1 без детализации каждого шага по рис.2. Последовательность шагов изменения роли не играет, каждая записанная версия ПП дает полный комплект всех сведений для генерации изображения. Возможен скачок через несколько шагов.

**Автосохранение.** Для борьбы с угрозой 2 применяется широко распространенный метод автоматического сохранения информации через заданный промежуток времени - автосохранение. Специфика комплексной САПР приводит к необходимости автосохранения не только чертежа, но и текущих параметрических представлений. С интервалом не менее заданного на диске автоматически создаются копии чертежа и ПП текущей задачи, предыдущая копия с диска удаляется. При нормальном завершении задачи или работы с чертежом автоматически созданные копии удаляются. В случае некорректного завершения работы (например, из-за отключения электропитания) эти копии остаются, и при следующей загрузке пользователю будет предложено восстановить чертеж и ПП задачи.

**Автоматическое резервное копирование.** Защита от случайной подмены файла чертежа обеспечивается путем автоматического резервного копирования всех файлов, подвергшихся изменению за период с предыдущего копирования, в отдельный каталог диска с именем, определяемым датой создания резервных копий. Пользователь назначает дни недели, в которые следует проводить автоматическое резервное копирование, и при первом запуске САПР в эти дни оно осуществляется. Забота об удалении ставших ненужными резервных копий лежит на пользователе.

**Электронная подпись.** Для сохранения и контроля целостности чертежа (борьба с угрозой 4) применяется метод электронной подписи. Он состоит в том, что при наличии хотя бы одной подписи запрещено любое изменение чертежа. Подписи защищены личными паролями. Наличие подписи на чертеже гарантирует совпадение его содержания с тем, что было в момент подписания. Снять подпись может любой пользователь, после чего чертеж можно изменять. Но при этом будет видно, что чертеж не подписан.

**Заключение.** Описанные четыре угрозы информации и методы ее защиты в комплексной САПР реконструкции предприятий являются в значительной степени общими и для других программных средств редактирования пользовательских файлов, поскольку состав угроз определен практическими потребностями. Индивидуальные особенности реализация ЗИ продиктованы прежде всего комплексностью САПР, поддерживающей несколько информационных моделей объектов проектирования.

# БИБЛИОГРАФИЧЕСКИЙ СПИСОК